\documentclass[
 ,draft            
  ]
  {aipproc}

\layoutstyle{8x11double}

\begin{document}

\title{A strict lower-limit EBL \\Applications on gamma-ray absorption}

\author{Tanja M. Kneiske}{
  address={Institute f\"ur Experimentalphysik, University of Hamburg, Luruper Chaussee 149, D-22761 Hamburg, Germany}
}

\author{Herve Dole}{
  address={Institut d'Astrophysique Spatiale, Universit\'e Paris Sud 11 \& CNRS, b\^at 121, F-91405 Orsay, France}
}


\begin{abstract}
A strict lower limit flux for the extragalactic background light from ultraviolet
to the far-infrared photon energies is presented. The spectral energy distribution is derived using 
an established EBL model based on galaxy formation. The model parameters are chosen to fit the lower limit
data from number count observations in particular recent results by the SPITZER infrared space telescope.
A lower limit EBL model is needed to calculate guaranteed absorption due to pair production in extragalactic  
gamma-ray sources as in TeV blazars.
\end{abstract}

\maketitle


\section{Introduction}
The extragalactic diffuse radiation, also called
extragalactic background light (EBL), is the relic emission of galaxy
formation and evolution, and is produced by direct star light (UV and
visible ranges) and light reprocessed by the interstellar dust
(infrared to submillimeter ranges).	

The EBL is difficult
to measure directly because of strong foreground contamination. Thus,
upper limits have been derived by observing the isotropic emission
component \cite{hauser_cosmic_2001}.  Another
method consists in using integrated galaxy number counts which has
been improved during the last years by sensitive telescopes like
{\it Spitzer}, to get lower limits.
To overcome the poor
constraints at far-infrared wavelengths, a stacking analysis of near-
and mid-infrared sources is used \cite{dole_cosmic_2006}, to
significantly resolve the cosmic infrared background, leading to
constraining lower limits.

In addition to observational constraints, models are being
developed. The main contribution to the infrared diffuse radiation is
produced in stars and galaxies too far away for being detected as
distinct sources.  The physics of stars, the composition and spatial
distribution of the interstellar medium can be estimated for close-by
galaxies but not easily generalized on global scales. Thus, different
formalisms may result in discrepancies of the EBL flux at the order of
5 to 10 at certain wavelengths, despite an overall reasonable shape
(for a detailed discussion see \cite{hauser_cosmic_2001}).  
The model used in this paper is an
updated version discussed in \cite{kneiske_implications_2002} and 
\cite{kneiske_implications_2004}.

Beside the study of stellar populations on global scales, another
effect triggered a great interest in the EBL flux: high-energy gamma
rays traveling through intergalactic space can produce
electron-positron pairs in collisions with low energy photons from the
extragalactic background light \cite{gould_opacity_1966}.  Despite this effect,
Cherenkov telescopes have been discovered a great number of
extragalactic high energy gamma-ray sources at unexpected large
redshift e.g. \cite{the_magic_collaboration_very-high-energy_2008}.
The H.E.S.S. collaboration derived an upper limit for the EBL between 1
and 4 micron, which is very close to the optical number counts by the
{\it Hubble Space Telescope} \cite{madau_deep_2000}.
The problem is that the so called ''upper limit'' strongly
depends on the assumption of the intrinsic blazar spectrum. 

Contrary to this upper limit, a lower limit is derived \cite{kneiske_strict_2008}.
Fitting an EBL model to lower limit data from optical to infrared
energies leads to a strict lower limit for the extragalactic
background light.  It will provide a minimum correction for
extragalactic gamma-ray sources due to photon photon pair-production.
In the next section the EBL model and data are summarized briefly. 
In the sections to follow the absorption of high energy gamma-rays is
studied using the lower limit model. Observed high energy gamma-ray sources
are shown with their possible intrinsic GeV spectra.
Throughout this paper, we adopt a
cosmology with $h=0.72$, $\Omega_{\rm M} = 0.3$ and $\Omega_{\Lambda} =
0.7$.
\begin{figure}
  \includegraphics[height=.5\textheight]{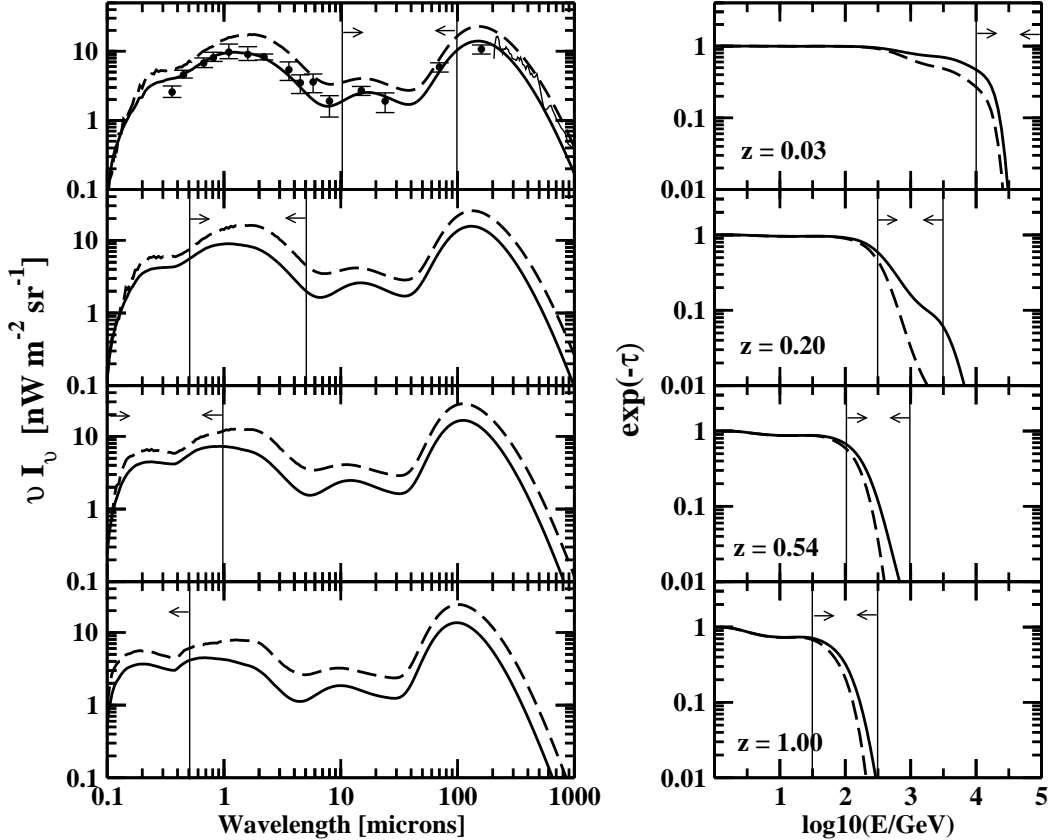}
  
  \caption{Comoving flux of the extragalactic background light and corresponding extinction of gamma-rays
  at four different redshifts.
  The solid line represents the lower-limit EBL introduced here while the dashed line
  is the old ''best-fit'' model described in \cite{kneiske_implications_2004}. Meaning of arrows
  and thin solid lines see text.}
\end{figure}

\section{Model and Observations}
The review of the lower limits on the Extragalactic Background Light
measurements are presented in detail in \cite{kneiske_strict_2008}.
Most of them come from the integration of number counts,
not from direct measurements of surface brightness, subject to strong
foreground emission contamination.  This method is based on the simple
counting of detected galaxies on a given sky area of a deep survey, a
completeness correction, and the integration of the number counts: it
is usually robust and gives lower limits; variance due to large-scale
structure may affect the results, and are usually taken into account
in the error bars.
Any model of the EBL should thus lie above these observed
limits. 

The EBL model has been first described in \cite{kneiske_implications_2002}.  The
parameter range has been studied and it has been used for correcting
spectra of extragalactic gamma-ray sources in \cite{kneiske_implications_2004}.
The main idea is to distinguish between two different star forming
regions in the universe which depends on the the amount of dust
present.
Two steps have been performed to calculate the EBL.  First the EBL has
been derived up to a redshift of two. This takes into account that
data from number counts are only able to resolve galaxies up to a
certain redshift which depends on the flux limit of the instrument. 
Note that the redshift
for the emissivity does not change to account for stars which have
formed at redshifts $z>2$ and are still present in galaxies at $z<2$.
The EBL is fitted to the data shown in the top panel of Fig.1.  The parameters
have been chosen by using a minimum $\chi^2$ fitting routine.  As a
second step the total EBL is calculated up to a redshift of $z=5$.
The main parameters are shown in table 1 in \cite{kneiske_strict_2008}.  

The result is shown as solid line in the left panel of Fig.1. The comoving flux of the 
EBL is shown at four different redshifts (0, 0.2, 0.54 and 1.0). 
At low redshift the optical and infrared peak are almost on the same level while for
higher redshifts the infrared peak dominates. 
The dashed line is the flux from  
the "best-fit" model described in \cite{kneiske_implications_2004} which is a factor
of about 2 higher.

\section{Gamma-Ray Absorption}
The optical depth $\tau_{\gamma \gamma}$ for gamma-rays in the universe can now be calculated following the
equation in \cite{kneiske_implications_2004}. The effect of absorption for sources
at different redshifts $z$ is shown in the right panel of Fig. 1.
Here the extinction $\exp(-\tau_{\gamma \gamma})$ is plotted as a function of gamma-ray
energy. A value of one means that the photons can pass aneffected through 
extragalactic space. A smaller value
leads to higher absorption effects. The effect is often described in terms
of the so called ''cut-off energy'' $E_c$ which is defined as $\tau_{\gamma \gamma}(E_c)=1$.
Photons with energies $E>E_c$ are absorbed by more than a factor of $1/e$. For those
photons the universe is optically thick.
The cut-off energy depends on the redshift of the source and is about 12~TeV, 
500~GeV, 185~GeV and 100~GeV for$z=0.03, 0.2, 0.54$ and 1.0 respectively.
Due to the peaked cross section for this process the absorption mainly 
takes place between EBL photons with energy $\epsilon{\mathrm EBL}$
and gamma-ray photons with energies $E_\gamma$ if $\epsilon_{\mathrm EBL} E_\gamma
\leq 2m_e c^2 \propto 1.2$~TeV$^2$. The thin solid lines and arrows
in Fig.1 are bracketing the corresponding gamma-ray and EBL energies depending
on redshift. For example at a redshift of $0.03$ the absorption 
effects mainly the photons with energies 100~TeV$> E_\gamma > 10$~TeV 
while interacting with EBL photons with $10 > \lambda > 100 \mu$m.

As an example for photon-photon pair production a sample of blazars have been studied.
In Fig.2 the flux of sixteen blazars observed with different Cherenkov Telescopes
like (Whipple, Hegra, MAGIC and H.E.S.S.)
is plotted at GeV energies. The spectra have been scaled by numbers between
$10^{10}$ and $10^{25}$.
The lower limit EBL model has been used to correct the intrinsic 
spectra which can now be compared with the observed data (solid lines in Fig.2).
An intrinsic power-law spectrum has been assumed for each detected source. 
The spectral indices $\alpha$ are shown in Fig.2 as first number
in the legend. The second number is the redshift $z_s$ of the source.
The following relation has been used 
\begin{equation}
\frac{dN}{dE}\propto E^{\alpha}\ \exp(-\tau_{\gamma \gamma}(E_\gamma,z_s, n_{EBL}))
\end{equation}
Note that the power-law index has been chosen without looking into 
statistical details. They are not representing best-fit values but
are rather crude estimates to show the absorption effect with the lower limit
EBL. The spectral indices and therefore the intrinsic blazar
properties are quite different.
A more detailed analysis is only useful if the results can be compared
with a detailed theoretical model for each blazar separately.

\begin{figure}
  \includegraphics[height=.53\textheight]{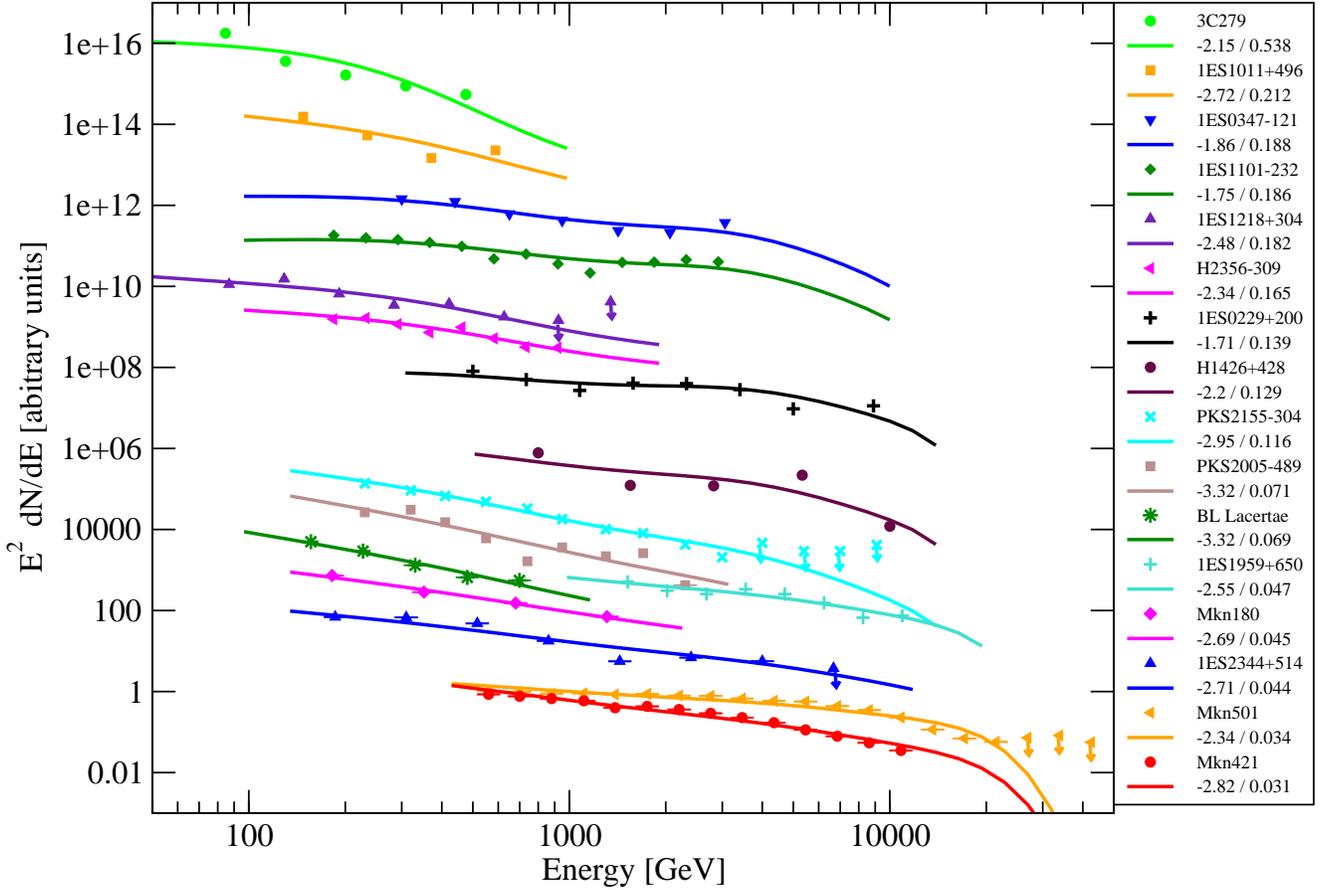}
  \caption{
  Observed spectral energy
distributions for blazars indicated in the figure. The
sources are ordered by their redshift, from high
(top) to low redshift (bottom). The total flux is normalized
for a better visualization. The lines are model spectra
corrected for minimum EBL absorption, described in the text.
Numbers on the right indicate the spectra index $\alpha$ and
the redshift of the source.
  }
\end{figure}

\section{Conclusions}
In this proceeding we have discussed the effect of pair production on
extragalactic gamma-ray sources using a lower limit EBL model based on
strict lower limit coming from infrared number counts.  The lower-limit
EBL is still in agreement with upper limits derived so far from the
process of pair production with very high energy gamma-ray emission by
BL Lacs.  

The intrinsic power-law spectra derived with the lower-limit EBL are
still in agreement with standard accelaration mechanisms proposed for
relativistic jets in blazars.  We find, however, that using a simple
power-law to fit the observed spectra corrected for absorption is not
optimal, and intrinsic spectra exhibit large variations between
sources.  For example Mkn501 and Mkn421 would be better fit by a
power-law with exponential cut-off. Furthermore, low photon statistics
limits the accuracy at which individual intrinsic spectra can be
recovered; Gamma-ray spectra with higher statistics, like we were able
to observe from PKS2155-304 in its extraordinary flair 
\cite{aharonian_exceptional_2007}, are needed for a detailed studied of each single blazar.
Finally, there might be a selection effect with redshift towards lower
spectral indices, that is so far not estimated.

Despite the use of the power-law approximation for the intrinsic
spectra, the modeled spectra corrected for EBL absorption using our
lower-limit EBL reproduce qualtiatively well the observations
(Fig. 2). This was not particulary expected, given for instance the
unrealisticly low cosmic star formation rate that our lower-limit EBL
implies \cite{kneiske_strict_2008}.

Another hint towards more complex blazar models would be
if future EBL limits from GeV/TeV observations become
lower. If they drop below the strict lower-limit EBL presented here,
observations of galaxies and star formation are violated and the
standard assumptions on blazar physics have to be revised. Attempts have
already been made like more detailed SSC simulations \cite{stecker_blazar_2007}
 which are  only 
leading to very small
changes of the photon index in BL Lacs. Other solution 
might be intrinsic absorption \cite{donea_radiation_2003}, \cite{sitarek_modification_2008}
or even exotic particle processes which prevent the gamma-ray photons
from being absorbed \cite{de_angelis_axion-like_2008}.

Lower limits measurements and the lower-limit EBL model are available online\footnote{in
  Orsay: http://www.ias.u-psud.fr/irgalaxies/ and in Hamburg:
  http://www.desy.de/\~ kneiske}.





\bibliographystyle{aipproc}   

\bibliography{Heid08_EBL}

\IfFileExists{\jobname.bbl}{}
 {\typeout{}
  \typeout{******************************************}
  \typeout{** Please run "bibtex \jobname" to optain}
  \typeout{** the bibliography and then re-run LaTeX}
  \typeout{** twice to fix the references!}
  \typeout{******************************************}
  \typeout{}
 }


\end{document}